\documentclass[aps,prb,showpacs,superscriptaddress,twocolumn,reprint]{revtex4-1}
\usepackage{hyperref}
\usepackage{graphicx, epstopdf, braket, amsmath, amssymb, natbib, xcolor}

\pdfoutput=1

\renewcommand\vec{\mathbf}

\begin{document}
    \title{Semimetallic features in quantum transport through a gate-defined point contact in bilayer graphene}
    \author{T.L.M. Lane}
    \email{thomas.lane-3@postgrad.manchester.ac.uk}
    \affiliation{National Graphene Institute, University of Manchester, Manchester, M13 9PL, UK}
    \affiliation{School of Physics and Astronomy, University of Manchester, Manchester, M13 9PL, UK}
    \author{A. Knothe}
    \affiliation{National Graphene Institute, University of Manchester, Manchester, M13 9PL, UK}
    \author{V.I. Fal'ko}
    \affiliation{National Graphene Institute, University of Manchester, Manchester, M13 9PL, UK}
    \affiliation{School of Physics and Astronomy, University of Manchester, Manchester, M13 9PL, UK}

    \begin{abstract}
        We demonstrate that, at the onset of conduction, an electrostatically defined quantum wire in bilayer graphene (BLG) with an interlayer asymmetry gap may act as a 1D semimetal, due to the multiple minivalley dispersion of its lowest subband. Formation of a non-monotonic subband coincides with a near-degeneracy between the bottom edges of the lowest two subbands in the wire spectrum, suggesting an $8e^2/h$ step at the conduction threshold, and the semimetallic behaviour of the lowest subband in the wire would be manifest as resonance transmission peaks on an $8e^2/h$ conductance plateau.
    \end{abstract}

    \maketitle

\begin{figure}[!ht]
        \centering
        \includegraphics[width=\linewidth]{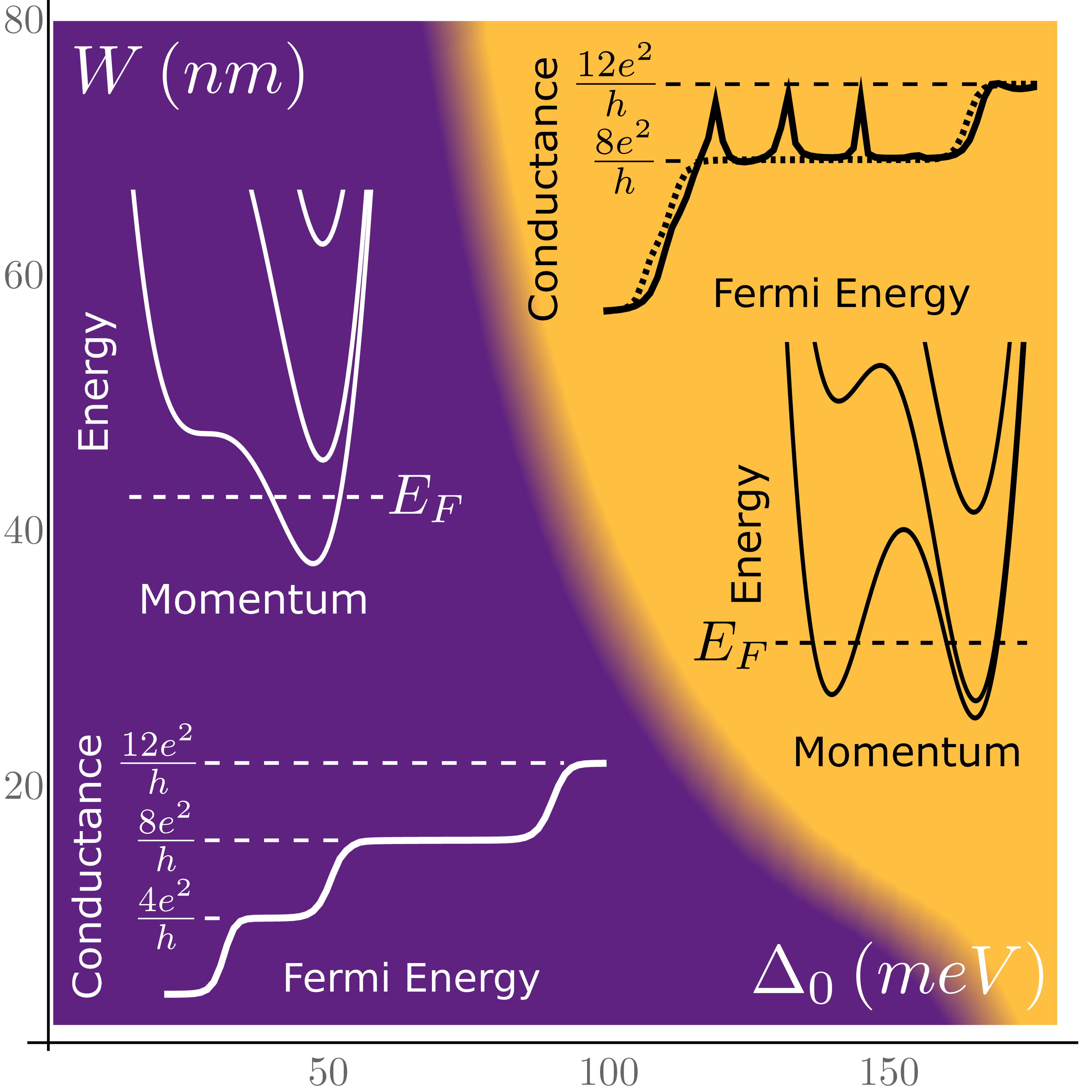}
        \caption{Parametric ranges for two quantum transport regimes in a BLG channel with zigzag orientation along the transport axis. For a small channel width, $W$, or small interlayer gap, $\Delta_0$, the subbands in the BLG wire are approximately quadratic and well-separated, leading to the regular ladder of $4e^2/h$ conduction steps (white insets). For a larger $\Delta_0$ or wider $W$, the lowest conduction subbands feature band inversions with multiple band minima / additional degeneracies (black insets). This enables coherent forwards scattering resonances, resulting in additional conductance spikes (solid black line) above an initial step of $8e^2/h$ specific to the adiabatic limit, as indicated by the dotted black line.}
        \label{fig:phase_diagram}
    \end{figure}

    Quantum transport in nanostructures is, very often, associated with the $e^2/h$ conductance quantisation in ballistic point contacts \cite{van1988quantized, van1989coherent, reznikov1995temporal, krans1995signature, qu2016quantized}. The onset of ballistic transport in the wires of conventional semiconductors, with a plain parabolic dispersion of carriers near the conduction or valence band edge, has been studied in detail \cite{buttiker1986role,beenakker1991quantum,datta1997electronic,sohn2013mesoscopic,elzerman2004single,petta2005coherent,Glazman1988}. Here, we study the onset of ballistic transport in a material where the band edge for the dispersion of charge carriers inside the wire has a non-monotonic form, like the one illustrated by the inset in the r.h.s. of Fig.~\ref{fig:phase_diagram}. An example of a system that displays such an unusual dispersion is electrostatically defined quantum wires in gapped bilayer graphene (BLG) \cite{Falko2007,Droscher2012,Hunt2017,Overweg2018,kraft2018,Hamer2018,Lee2018}, where, for some parameters, the lowest-energy electron subbands contain both `electron-like' and inverted (`hole-like') parts \cite{Knothe2018,Overweg2018a}, making the wire with low doping semimetallic. Such non-monotonic dispersion is inherited by electrons in a quantum wire from the three minivalleys at the band edges of gapped BLG \cite{McCann2006,varlet2014anomalous}, which are more pronounced in BLG with a larger interlayer asymmetry gap.

    Below, we show that inner branches of such non-monotonic dispersion, Fig.~\ref{fig:phase_diagram}, are responsible for the formation of confined states bouncing forth and back in the semimetallic segment of the wire, which appears to be the region featuring the highest energy of the lowest subband edge. In a wire with smooth (adiabatic) confinement edges, outer branches of the subband dispersion would propagate ballistically without interacting with these confined states. In a realistic wire geometry, the bouncing states associated with the inner semimetallic part of the non-monotonic electron subbands generate transmission resonances for the incoming higher-energy states that otherwise would not pass across the wire, on top of the quantised conductance plateaux due to the lowest subbands outer branches. We find that, as shown in Fig.~\ref{fig:phase_diagram}, this effect is more pronounced in wider quantum wires, or when a larger interlayer asymmetry gap is induced in BLG via electrostatic gating.

\begin{figure}[!ht]
        \centering
        \includegraphics[width=\linewidth]{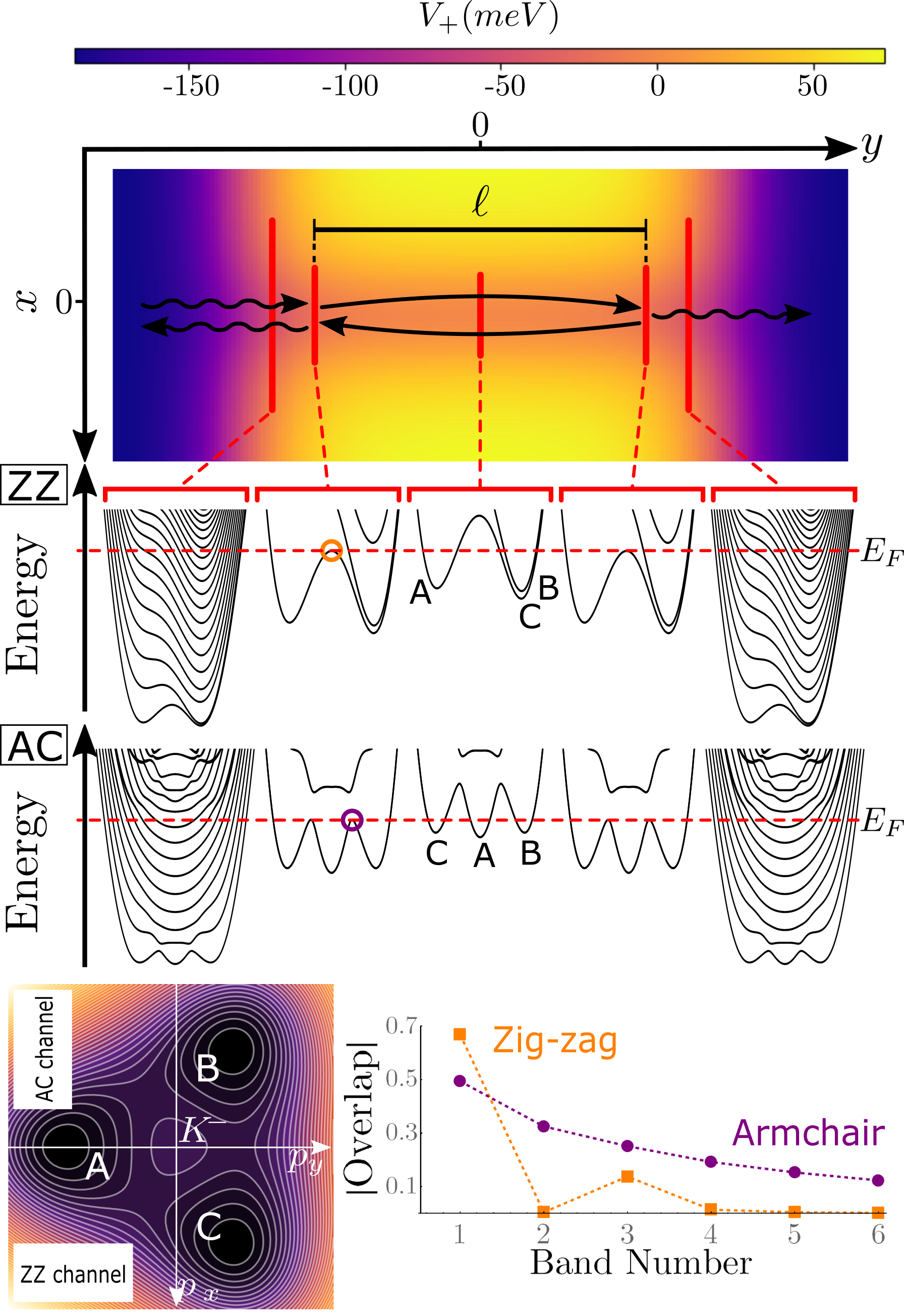}
        \caption{Top: Potential profile, $V_+(x,y)$, confining electrons to the upper layer in gapped BLG along a quasi-1D channel between two leads. Crosses on the profile illustrate the edges of the semimetallic segment of the wire which supports standing waves in the `hole-like' part of the non-monotonic lowest energy subbands, resulting in transmission resonances. Middle: 1D energy spectra (plotted vs travelling momentum $p_y$) for cuts along a zigzag (ZZ) and armchair (AC) oriented channel. Bottom left: Minivalleys A-C are identified in the contour plot illustrating the trigonally warped low energy band structure (Eq.\eqref{eq:dispersion}) of gapped BLG around the $K^-$ point. Bottom right: Overlap between subband states at the Fermi energy in the leads and the top of the inverted branch of the lowest energy subband at the edge of the semimetallic wire. Overlaps computed using the continuum model of gapped BLG in the presence of a confining potential and spatially modified gap (similar to the methods used in Ref.~\onlinecite{Knothe2018}).}
        \label{fig:channel}
    \end{figure}

    To model electronic transport of quantum wires in BLG, we implement a tight binding model for a channel between two electronic reservoirs (see top panel in Fig.~\ref{fig:channel}) described by the Hamiltonian,
    \begin{eqnarray}
        \hat{\mathcal{H}}&=&-\gamma_0\sum\limits_\eta\sum\limits_{\braket{i,j}}\left\{a_{\eta,i}^\dagger b_{\eta,j}^{} + H.c.\right\}\label{eq:hamiltonian}\\
        &&- \gamma_1\sum\limits_{\braket{i}}\{a_{+,i}^\dagger b_{-,i}^{} + H.c.\}
        -\gamma_3\sum\limits_{\braket{\braket{i,j}}}\{a_{-,i}^\dagger b_{+,j}^{} + H.c.\}\nonumber\\
        &&+ \sum\limits_\eta\sum\limits_i V_\eta(x_i,y_i)\left\{a_{\eta,i}^\dagger a_{\eta,i}^{} + b_{\eta,i}^\dagger b_{\eta,i}^{}\right\}.\nonumber
    \end{eqnarray}
    Here, $a_{\eta,i}^\dagger$/$b_{\eta,i}^\dagger$ ($a_{\eta,i}^{}$/$b_{\eta,i}^{}$) are creation (annihilation) operators for electrons at A/B sites in the upper ($\eta=+1$) and lower ($\eta=-1$) layer at position $\mathbf{r}_i = (x_i, y_i)$. Hopping parameters describe the intralayer nearest-neighbour coupling ($\gamma_0 = 3.16$~eV) the interlayer nearest-neighbour coupling ($\gamma_1 = 0.39$~eV) and skew coupling ($\gamma_3 = 0.38$~eV). The sum over $\braket{i,j}$ runs over all intralayer nearest neighbours, $\braket{i}$ runs over all dimer sites, $\braket{\braket{i,j}}$ runs over all nearest neighbour non-dimer sites, and $i$ runs over every atomic site. The hopping part of the Hamiltonian, together with the interlayer asymmetry gap, $\pm\frac{\Delta}{2}$, determines the low energy BLG spectrum in the $K^\xi$ valley,
    \begin{equation}
        \varepsilon_\text{low}^2=(v_3p)^2-\xi \frac{v_3p^3}{m}\cos(3\phi)+\frac{p^2}{2m}\left[\frac{p^2}{2m}-\frac{\Delta^2}{\gamma_1}\right]+\frac{\Delta^2}{4},\label{eq:dispersion}
    \end{equation}
    where $\mathbf{p}=p(\cos(\phi),\sin(\phi))$, $v=10^8$~ms$^{-1}$ is the Dirac velocity in graphene related to $\gamma_0$ and $m=\gamma_1/2v^2$ ($vp/\gamma_1\ll1$) and $v_3/v=\gamma_3/\gamma_0\ll1$. This dispersion is illustrated by the topographic map on the l.h.s. of the bottom panel of Fig.~\ref{fig:channel}.

    The conduction channel (quantum wire) is determined by a spatially modulated potential, $U(x) = U_0/\cosh(x/W)$, and interlayer asymmetry gap, $\Delta(x) = \Delta_0[1-\beta/\cosh(x/W)]$ as,
    \begin{eqnarray}
        &V_\eta(x,y) = \left[U(x)+\frac{\eta}{2}\Delta(x)\right]f(y)+\left[V_0 + \frac{\eta}{2}\delta\right]\left[1-f(y)\right],&\nonumber\\
        &f(y)=\frac{\sinh(L/\Lambda)}{\cosh(L/\Lambda)+\cosh(2y/\Lambda)}.\label{eq:potential}&
    \end{eqnarray}
    Here, $\Delta_0$ is the gap in the insulating area of BLG, $U_0$ and $\beta$ define the BLG band edges inside the one-dimensional channel, $\delta$ and $V_0$ are the interlayer energy gap and Fermi level within the lead region, and $W$ and $L$ define the width of the wire and its length, respectively. The adiabaticity parameter $\Lambda$ in the interpolation function $f(y)$ controls the smoothness of the wire. The profile of this confining potential is modelled after COMSOL simulations performed in support of experimental studies \cite{Overweg2018} and is consistent with the previous theoretical analysis of Ref.~\cite{Knothe2018}. The colour-map in the top panel of Fig.~\ref{fig:channel} sketches the profile of this confining potential for the upper layer of the bilayer, $V_+(x,y)$, illustrating the conduction channel (darker orange) squeezed between gapped BLG areas shown in bright yellow.

    Diagonalising the Hamiltonian at different cross sections (e.g., along the red lines indicated on the potential profile in Fig.~\ref{fig:channel}), we find the local spectrum of conduction bands for electrons in the wire. These spectra are illustrated in the middle panels of Fig.~\ref{fig:channel} for BLG with a large gap and for the channel oriented along the zigzag (ZZ) or armchair (AC) direction \footnote{Spectra are shown in the low energy regime around the $K^-$ valley, with spectra around the $K^+$ valley recovered by exchanging $p_x\rightarrow-p_x$.}. In both cases, the low energy bands in the narrowest part of the wire exhibit multiple band minima \cite{Knothe2018} that originate from the minivalleys A,B,C of gapped BLG, as marked on the lower left panel of Fig.~\ref{fig:channel}.

    To study the transport along such a wire we utilise a recursive non-equilibrium Green's function algorithm \cite{thouless1981conductivity,lewenkopf2013recursive} with minimal unit cells along the wire axis (see Appendix A). Semi-infinite leads are included via the self-energies of their surface Green's functions, evaluated using the Sancho-Rubio method \cite{sancho1985highly} (see Appendix B). In order to eliminate edge states arising from the finite width of the device region in the non-transport direction we apply periodic boundary conditions. Then, we calculate the transfer matrix by sampling Green's functions which describe atoms in the first channel up to the Fermi energy. Conductance of the wire is computed as,
    \begin{equation}
        G=\frac{2e^2}{h} Tr[\mathbf{G}^\dagger_D\mathbf{\Gamma}_R\mathbf{G}^{}_D\mathbf{\Gamma}_L]\label{eq:conductance},
    \end{equation}
    where the quantum unit $2e^2/h$ accounts for spin degeneracy, $\mathbf{G}^{}_D$ is the fully connected Green's function of the device region at the Fermi level, $E_F$, and $\mathbf{\Gamma}_{L/R}=-2Im[\mathbf{\Sigma}_{L/R}]$ accounts for the leads via their surface self energies, $\mathbf{\Sigma}_{L/R}$. Since we use the tight binding model Hamiltonian as a starting point, valley degeneracy of graphene is automatically recovered by the computation.

    \begin{figure}[!ht]
        \centering
        \includegraphics[width=\linewidth]{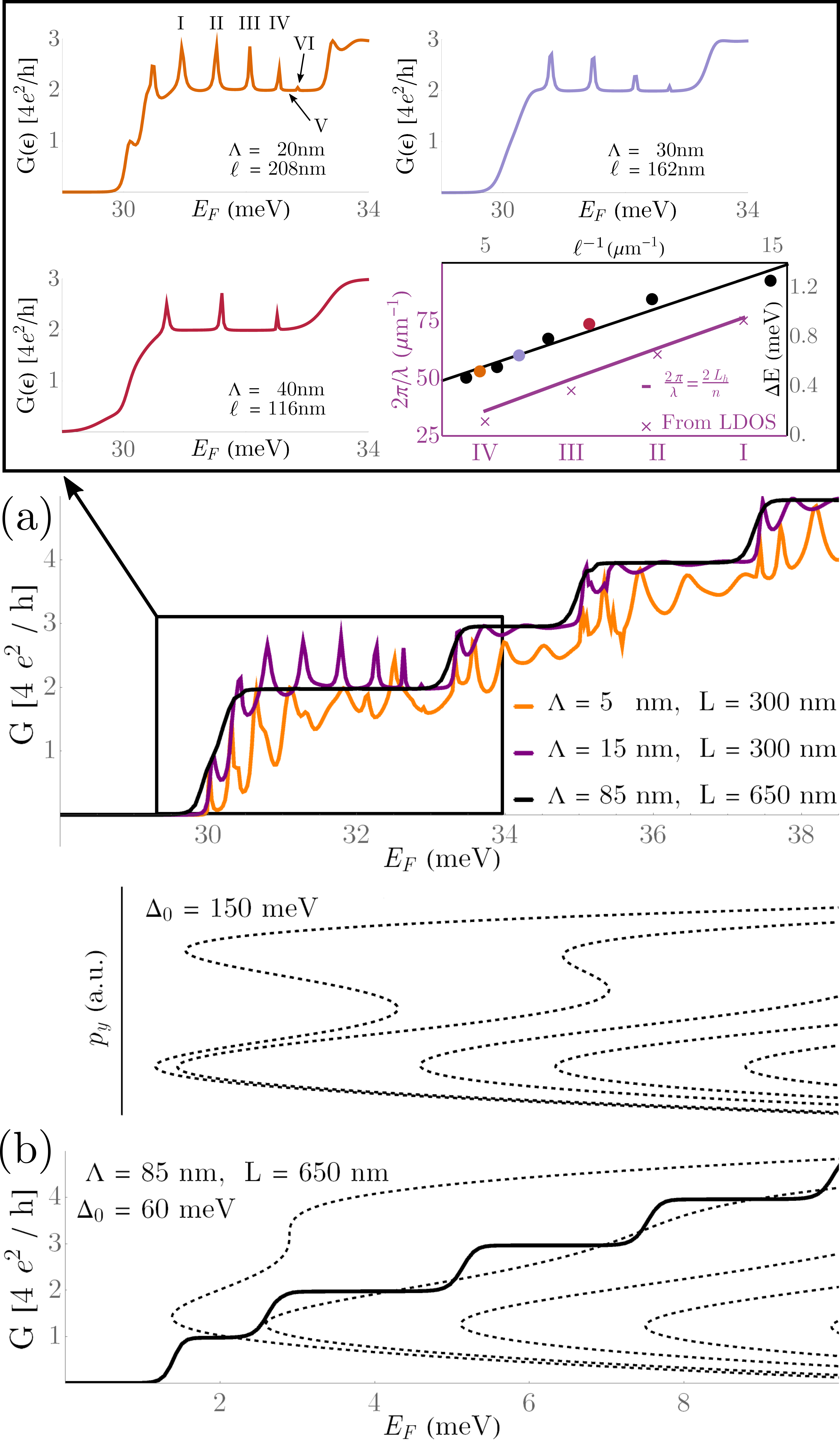}
        \caption{(a) Conductance, $G$, through a channel with zigzag lattice orientation for a range of smoothing distances, $\Lambda$. The corresponding channel spectra is illustrated below. Parameters used are $U_0=-20$~meV, $\Delta_0=150$~meV, $\beta=0.3$, $\delta=10$~meV, $V_0=-200$~meV, $L=300$~nm and $W=50$~nm. Inset above: The frequency of the resonance peaks on top of the first conductance plateau changes with the length of the semimetallic region, $\ell$. The average lateral energy spacing of the resonance peaks is proportional to the inverse of the effective channel length ($\Delta E\propto\ell^{-1}$), whilst the wavenumber, $k$, of standing waves within the channel varies linearly with channel length (crosses show data extracted from the LDOS images which are fit with a calculated curve assuming hard wall boundary conditions at the wire ends). (b) Transport characteristics for a weakly gapped system ($\Delta_0=60$~meV, other parameters the same as in (a)) which removes the non-monotonic features in the dispersion of the lead resulting in regular, adiabatic conductance steps of $4e^2/h$.}
        \label{fig:transport_zz}
    \end{figure}

    Figure~\ref{fig:transport_zz} illustrates the results of transport calculations for the ZZ oriented system for a range of smoothing lengths, $\Lambda$, with the corresponding energy spectra within the channel shown below. For an abrupt end of the wire (orange curve), there is significant noise and a decreased value of conductance. This arises from a combination of backscattering at the sharp potential boundary (comparable to the injected electron wavelength) and intersubband scattering. The black curve in the upper panel corresponds to an adiabatic channel with smoothing distance much longer than the incident electron wavelength. This conductance curve features an initial step of $8e^2/h$ (counting spin and valley degrees of freedom) followed by subsequent steps of $4e^2/h$. The origin of the double height of the first conductance step is related to the approximate degeneracy of the two lowest energy subband edges, which appears simultaneously with the formation of minivalleys in the subband dispersion \cite{Knothe2018}. In contrast, for an intermediate smoothing length (purple curve) we observe an $8e^2/h$ plateau with additional transmission resonances on top of it. These resonances appear only when the lowest energy subband features a band inversion. The point where the maximum of the inverted dispersion in the lowest subband crosses $E_F$ determines the turning points of the `hole-like' states of electrons (marked by crosses on the upper panel in Fig.~\ref{fig:channel}), creating standing waves within the semimetallic part of the wire which generate the transmission resonances. The efficiency of the resonances is determined by the scattering rates between incoming higher-energy branches in the wider part of the wire and those semimetallic states in the middle of it. A sharper profile of the wire promotes such scattering, whereas it is exponentially suppressed \cite{Glazman1988} in adiabatic wires ($\Lambda\rightarrow\infty$).

    \begin{figure*}[!ht]
        \centering
        \includegraphics[width=\linewidth]{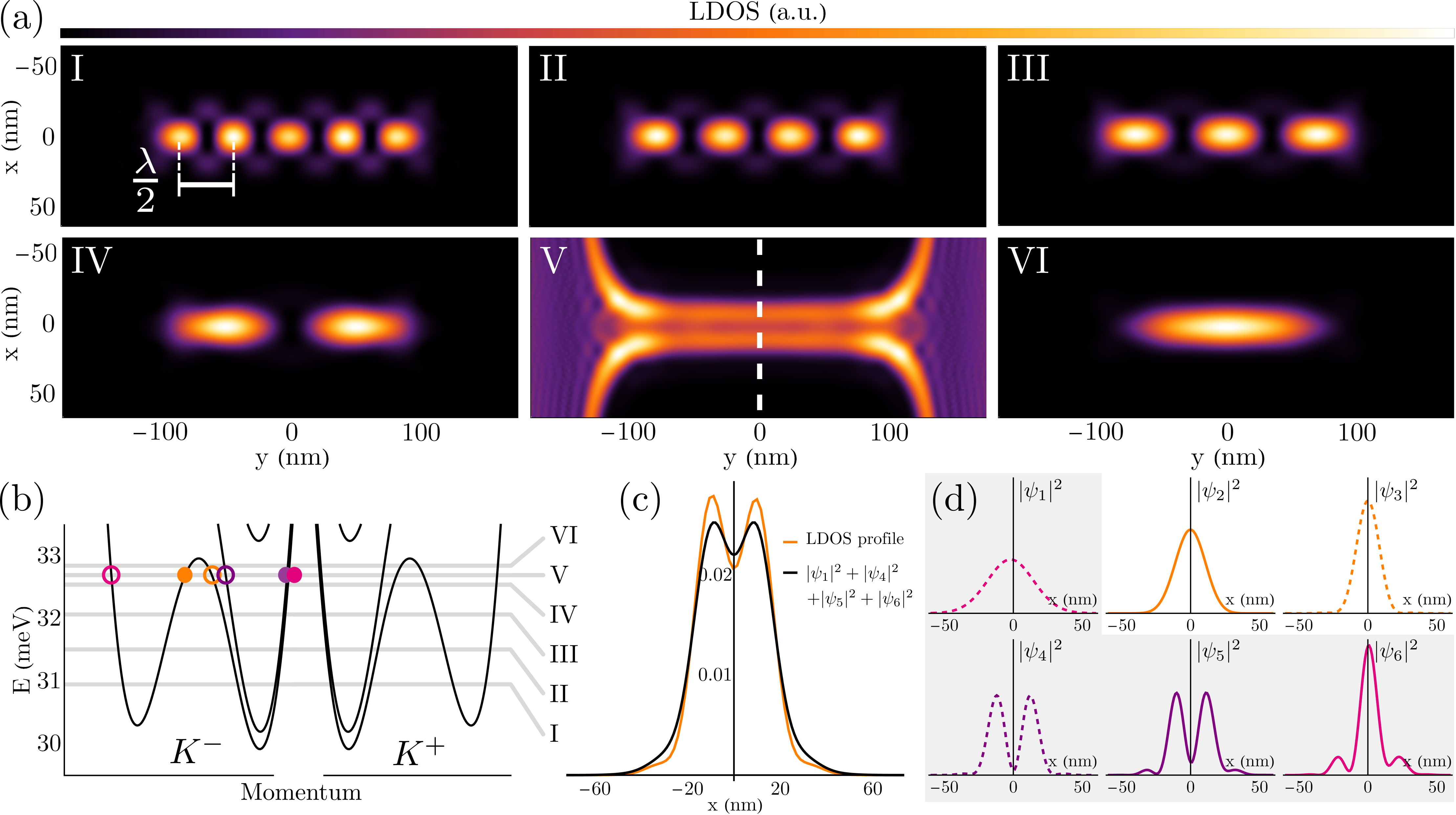}
        \caption{(a) Momentum-averaged local density of states (LDOS) at energy cuts along the first conductance plateau for a zigzag oriented wire. Device parameters are the same as in Fig.~\ref{fig:transport_zz}(a), with $\Lambda=20$~nm. Numerals I-VI identify the energy of each LDOS image in the channel's low energy subband spectra shown in panel (b), and in the corresponding conductance profile in the inset of Fig.~\ref{fig:transport_zz}(a). LDOS image V is for an energy on the first plateau (between peaks), whilst the other cuts correspond to the energies of resonance peaks. (c) The sum of `electron-like' states in panel (d) (black curve) as compared to the renormalised profile extracted from LDOS image V along the dashed white path. (d) Continuum wavefunctions calculated for the energy used in LDOS image V. Solid (dashed) wavefunctions are right (left) moving states corresponding to filled (empty) circles in the $K^-$ valley subband spectra in panel (b).}
        \label{fig:LDOS_analysis}
    \end{figure*}

    We calculate local density of states (LDOS) cuts across the energy range of the first conductance plateau (Fig.~\ref{fig:LDOS_analysis}). LDOS images I-IV and VI correspond to the energies of resonance peaks in the conductance profile\footnote{Standing waves are isolated from other quantum wire states by subtracting the LDOS averaged either side of the peak from the LDOS evaluated precisely at the peak.}, illustrated in the upper left inset of Fig.~\ref{fig:transport_zz}(a) (also identified in the $K^\pm$ valley subband spectra depicted in Fig.~\ref{fig:LDOS_analysis}(b)). The wavenumber of standing waves supported by the `hole-like' part of the wire varies linearly with the energy of the resonance peaks (see inset of Fig.~\ref{fig:transport_zz}(a)). Consecutively higher energy peaks are generated by increasingly low frequency standing waves within the channel, with the final resonance peak being consonant with the ground state standing wave which has a period of approximately twice the effective channel length.

    Figure \ref{fig:LDOS_analysis}(a) V depicts the LDOS for an energy lying between two resonance peaks, where conductance through the channel is at the $8e^2/h$ adiabatic limit. In this case, states are occupied with an approximately constant value along the length of the channel and exhibit a symmetric, double-peaked profile perpendicular to the transport axis. Extended states coming from the leads are shown less brightly due the much larger normalisation region of their support. In Fig.~\ref{fig:LDOS_analysis}(d) we show the continuum wavefunctions evaluated in the centre of the BLG wire for the same energy as used in LDOS image V and identified as coloured circles in Fig.~\ref{fig:LDOS_analysis}(b). Wavefunctions in the opposite valley can be recovered by taking $k\rightarrow-k$ and $x\rightarrow-x$. States in the inverted part of the spectrum (orange) are the symmetric states which give rise to the standing wave features previously discussed. Symmetric and antisymmetric states in the outer branches of the two lowest energy subbands are shown in pink and purple respectively. The latter of these two is the source of the `train-track' features in the plateau LDOS. Simply summing the contribution from these four states in each valley (without any free fit parameters) produces a symmetrical lateral profile which closely matches the profile along the dashed white path in LDOS image V (Fig.~\ref{fig:LDOS_analysis}(c)).

    We also calculate conductance through a wire for a reduced interlayer asymmetry gap with the same values of $L$ and $\Lambda$ (Fig.~\ref{fig:transport_zz}(b)). In this case, BLG wire subbands are approximately quadratic throughout the low energy range. Then, we find that the resonant peaks in conductance vanish, with the resulting conductance making a ladder of $4e^2/h$ steps. The parametric ranges for each of the two regimes described above are shown in Fig.~\ref{fig:phase_diagram}.

    \begin{figure}[!ht]
        \centering
        \includegraphics[width=\linewidth]{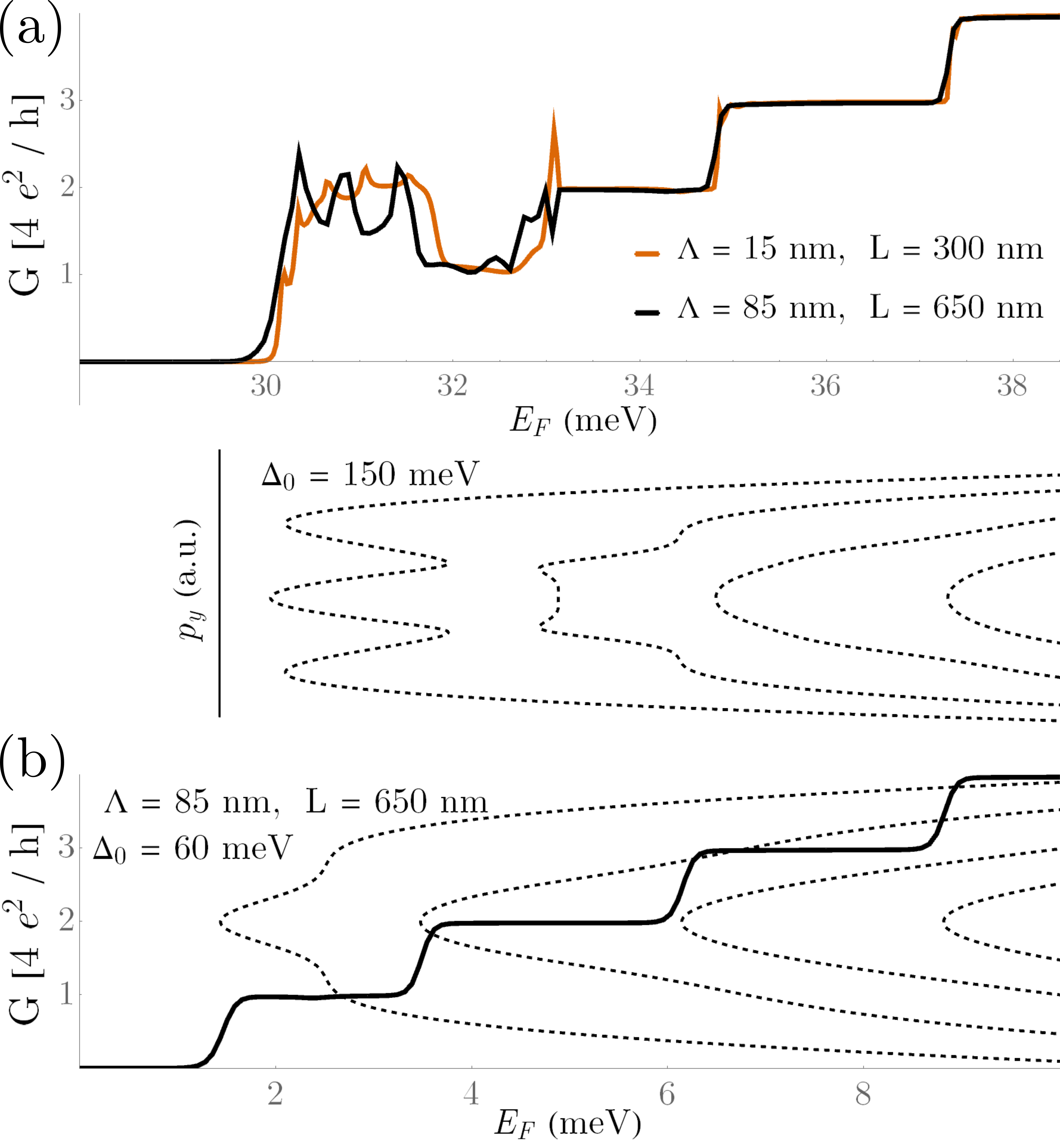}
        \caption{(a) Conductance, $G$, through a channel with armchair lattice orientation for a range of smoothing distances, $\Lambda$. The spectra within the channel is demonstrated below. The parameters used are the same as those in Fig.~\ref{fig:transport_zz}(a). (b) Transport characteristics for a weakly gapped channel with armchair orientation (parameters the same as in Fig.~\ref{fig:transport_zz}(b)). This again removes the band inversions, resulting in an approximately quadratic energy spectrum and a regular series of $4e^2/h$ conductance steps.}
        \label{fig:transport_ac}
    \end{figure}

    Conductance of a channel oriented along the armchair (AC) direction in graphene for both sharp (orange curve) and very smooth (black curve) channel profiles is shown in Fig.~\ref{fig:transport_ac}(a). As for ZZ wires, here, we find an initial step of $\approx8e^2/h$. However, this one is followed by a drop to a $4e^2/h$ value whilst the Fermi level lies between the energies of the lowest subband maxima and the bottom of the next subband in the narrowest part of the wire. Above that energy we recover the usual series of $4e^2/h$ conductance steps as we cross subsequent, approximately quadratic, band edges. Furthermore, we find that resonances on top of the first conductance plateau of $8e^2/h$ persist here for much smoother wire profiles (black curve), as compared to the ZZ oriented wires, whereas for the wires with a small interlayer asymmetry gap AC and ZZ cases produce very similar results with a simple $N\times4e^2/h$ conductance quantisation staircase (Fig.~\ref{fig:transport_ac}(b)). We believe that such difference is caused by the disparate symmetries between subband wavefunctions in the ZZ and AC orientated channels, leading to different selection rules for the intersubband scattering in the vicinity of the turning points for the `hole-like' states in the semimetallic middle part of the wire. We demonstrate this difference in Fig.~\ref{fig:channel}(lower r.h.s. panel), where we show the projections of $y$-dependent factors of the higher subband wavefunctions outside the constriction onto the wavefunction at the top of the inverted branch of the subband dispersion. In the case of a ZZ channel we observe zero overlap for even numbered bands and a rapid reduction in the magnitude of the overlap upon increasing the band index. For an AC channel the overlap obeys no selection rule and changes slowly with increasing band number, so that the resonances caused by the `hole-like' standing waves in the semimetallic part of the wire persist even in very smooth BLG constrictions.

    To conclude, we have noted that a sufficiently wide electrostatically defined wire in bilayer graphene with a substantial interlayer asymmetry gap contains a region which exhibits semimetallic properties. This feature has also been found to coincide with an additional near-degeneracy of the two lowest subbands at the BLG wire band edge. In this semimetallic segment of the wire the electron dispersion in the lowest subband is non-monotonic, generating standing waves for the `hole-like' states in the inverted part of the subband dispersion, bouncing between turning points at the wire ends. These peculiar standing wave states produce transmission resonances on the top of conductance plateaux (here, $8e^2/h$) specific for the adiabatic point contact. Decreasing the gap in bilayer graphene or narrowing the confinement potential in the wire removes the non-monotonic feature in the subband dispersion, restoring the usual staircase of $4e^2/h$ steps expected for an adiabatic contact in a material with a $4$-fold degeneracy in the spectrum.

    \begin{acknowledgments}
        The author would like to thank T. Aktor, A-P Jauho, S. Magorrian, K. Ensslin, C. Stampfer and R. Danneau for useful discussions. This work was funded by the Manchester Graphene-NOWNANO CDT EP/L-1548X, ERC Synergy Grant Hetero2D, and the European Graphene Flagship Project.
    \end{acknowledgments}

    \appendix

    \section{\label{A}Recursive Green's function methods}

\begin{figure}[!ht]
        \centering
        \includegraphics[width=\linewidth]{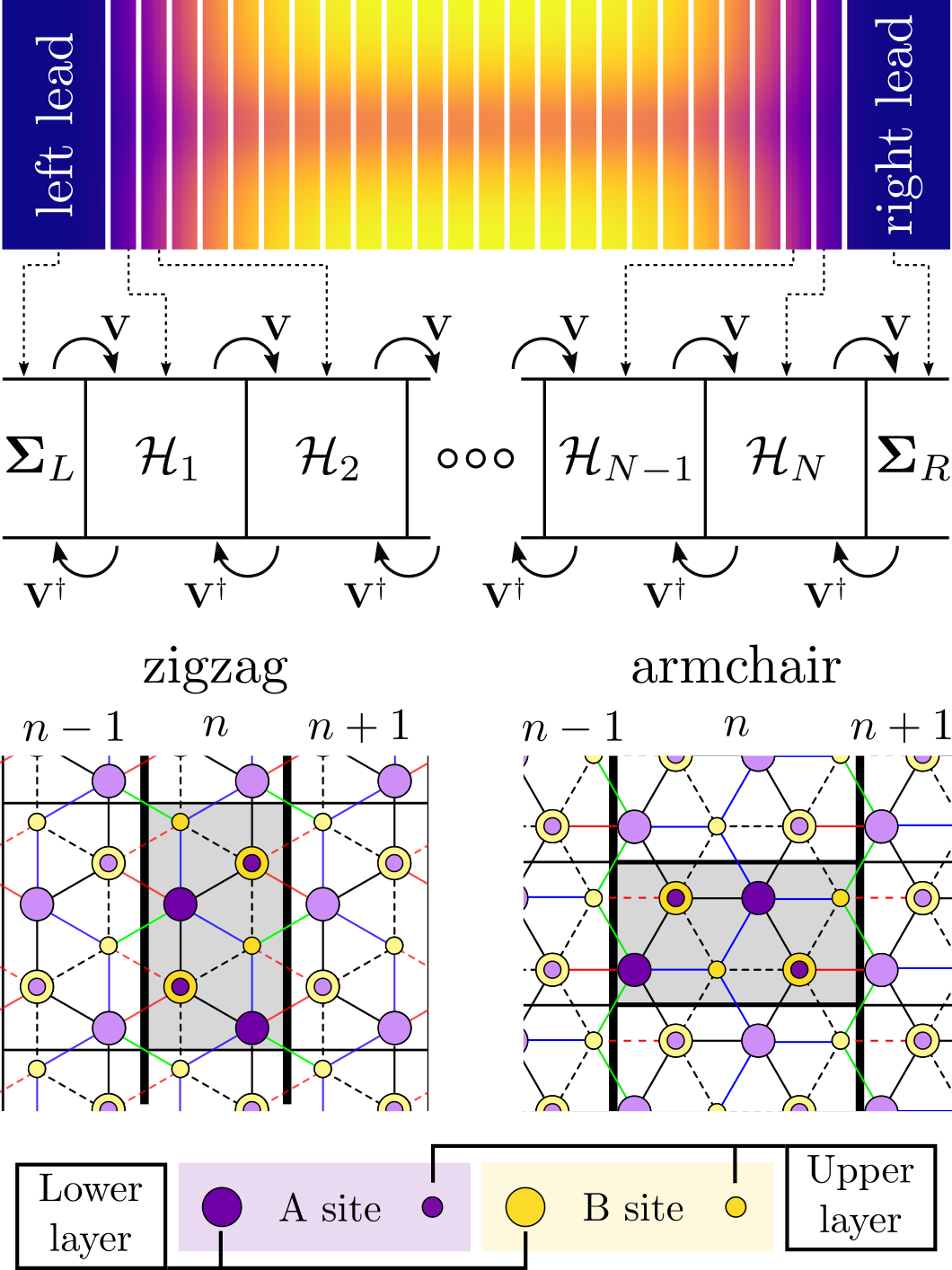}
        \caption{Top: Schematic of how the device (represented by the potential profile) is split up into slices in order to implement the recursive algorithm. Bottom: Sketch of the atomic arrangement in each slice for both zigzag and armchair orientation along the transport axis. Slices are constructed by tiling a minimal rectangular unit cell (e.g. grey zone) containing 8 atoms along the non-transport (vertical) axis. Intralayer ($\gamma_0$) hops are shown as solid (dashed) lines for the lower (upper) layer. Intralayer hops within a slice are shown in black, whilst hops between slices are shown in red. Skew interlayer ($\gamma_3$) hops within the slice are shown in blue and those between slices are shown in green. The thick black vertical lines define the edges of the $n$-\textit{th} slice.}
        \label{fig:cells}
    \end{figure}

    The recursive algorithm employed to calculate the conductance characteristics of the quantum wire follows the process outlined in Ref.~\onlinecite{lewenkopf2013recursive} for a two-probe system. The system is split into three components; the device region and the two lead regions. The device region is further broken up into narrow, independent slices connected in a chain along the transport axis as demonstrated in the upper section of Fig.~\ref{fig:cells}.
    Unperturbed retarded Green's functions for the independent cells, $\vec{g}$, and retarded Green's functions connected to the rest of the system, $\vec{G}$, are recursively related via the Dyson equation,
    \begin{equation}
        \vec{G}=\vec{g}+\vec{g}\vec{V}\vec{G},\label{eq:dyson}
    \end{equation}
    where $\vec{V}$ characterises the perturbation. Refactoring this expression for the perturbed Green's function we produce,
    \begin{equation}
        \vec{G}=(\vec{g}^{-1}-\vec{\Sigma})^{-1},\label{eq:dyson_02}
    \end{equation}
    where $\vec{g}^{-1}=(\varepsilon+i\zeta)\vec{I}-\vec{\mathcal{H}}$ is the unperturbed Green's function expressed in the energy domain with real-valued convergence factor $\zeta>0$, and $\vec{\Sigma}=\vec{V}\vec{g}\vec{V}^\dagger$ is the self energy added to the unperturbed Hamiltonian when connecting the decoupled systems. Parameter $\zeta$ ensures that the correct contour in the complex plane is selected when Fourier transforming the retarded Green's function from the time domain to the energy domain and ensures convergence of the ensuing calculation \cite{datta1997electronic,fetter2012quantum}.

    The two semi-infinite leads are described via surface self-energies at their interface with the device region (see Appendix~\ref{B}). Starting from one side of the device (i.e at the $n=1$ site), we express the Green's function of the first cell in terms of its unperturbed Hamiltonian, $\vec{\mathcal{H}}_1$, and the surface self energy of the connected lead, $\vec{\Sigma}_L$, as per Eq.~\eqref{eq:dyson_02}. We iterate across the device, calculating partially-connected Green's functions for each cell, $\vec{G}_n$. These are coupled back to the initial lead by including the self energy of the preceding, connected slice, $\vec{\Sigma}_{n-1}=\vec{V}\vec{g}_{n-1}\vec{V}^\dagger$. The opposite lead is attached by including its surface self energy in the perturbed Green's function of the final ($n=N$) cell, in addition to the self energy of the preceding ($n=N-1$) partially connected cell in the iteration. This final cell is now fully connected across the device to both the left and right leads and is sufficient to calculate conductance characteristics. States localised at the edges of the device region are removed by imposing periodic boundary conditions. We eliminate conductance contributions from parallel channels by including only Green's functions localised in the first channel in our conductance calculation. Further analysis, such as calculating the LDOS (local density of states, as in Fig.~\ref{fig:LDOS_analysis}), requires Green's functions for each cell to be fully connected to both leads, making a reverse algorithm that iterates in the opposite direction a necessity. The full recurrence relations for the left moving and right moving algorithm may be found in Ref.~\onlinecite{lewenkopf2013recursive}.

    The equation central to recursive process is Eq.~\eqref{eq:dyson_02}. The most computationally expensive aspect is the matrix inversion, which scales as the third power of the system size (total number of orbitals described by local Hamiltonian $\vec{\mathcal{H}}$). A single inversion is required per iteration of the recursive algorithm. By dividing the device region into $N$ slices of $M$ orbitals we reduce the scaling of the calculation by a factor $N^2$; from $\mathcal{O}([N\times M]^3)$ for a single cell spanning the entire device region to $\mathcal{O}(N\times M^3)$. To accommodate the smoothing regions between the lead and device sections of the channel the lattice is multiple times longer along the transport axis than it is wide (up to $5\times$ longer). Therefore, we optimise efficiency by minimising the number of orbitals per slice, $M$, whilst maximising the number of slices, $N$.

    The slices are constructed by tiling copies of the minimal rectangular unit cell (grey region containing 8 atoms in the lower panels of Fig.~\ref{fig:cells}) in the direction perpendicular to transport. The number of minimal cells along this axis is chosen such that the resulting lattice spans across the entire width of the channel. Since the minimal unit cell is not square, the number of tiles required to span the channel depends on lattice orientation. For a zigzag (armchair) orientated channel with $W=50$~nm the minimal unit cell is tiled 900 (1400) times, resulting in $M=7200$ ($M=11200$) atoms per slice. This results in an large orbital basis for each slice, making the calculation computationally taxing. For this reason we employ a scalable model of the graphene lattice as outlined in Refs.~\onlinecite{liu2015scalable,beconcini2016scaling}. This reduces the number of atoms required to span the channel by making the lattice constant, $a$, (and therefore the minimal unit cell) a factor $\sigma$ larger whilst scaling the hopping energies $\gamma_0$/$\gamma_3$ such that band velocities $v=\sqrt{3}a\gamma_0/2\hbar$ and $v_3=\sqrt{3}a\gamma_3/2\hbar$ remain fixed. Throughout this work $\sigma=8$.

    \section{\label{B}Lead description}

    The semi-infinite leads attached to either end of the channel are constructed of the same slices as in the rest of the device (see Fig.~\ref{fig:cells}). They are described using the Rubio-Sancho method given in Ref.~\onlinecite{sancho1985highly}. This is a highly efficient method which reduces an infinite series of identical slices to a single matrix of self energies, $\Sigma_{L/R}$, describing the surface of a left/right lead. Convergence of this method requires that all slices in each lead have the same structure and energy profile.

\begin{figure}[!ht]
        \centering
        \includegraphics[width=\linewidth]{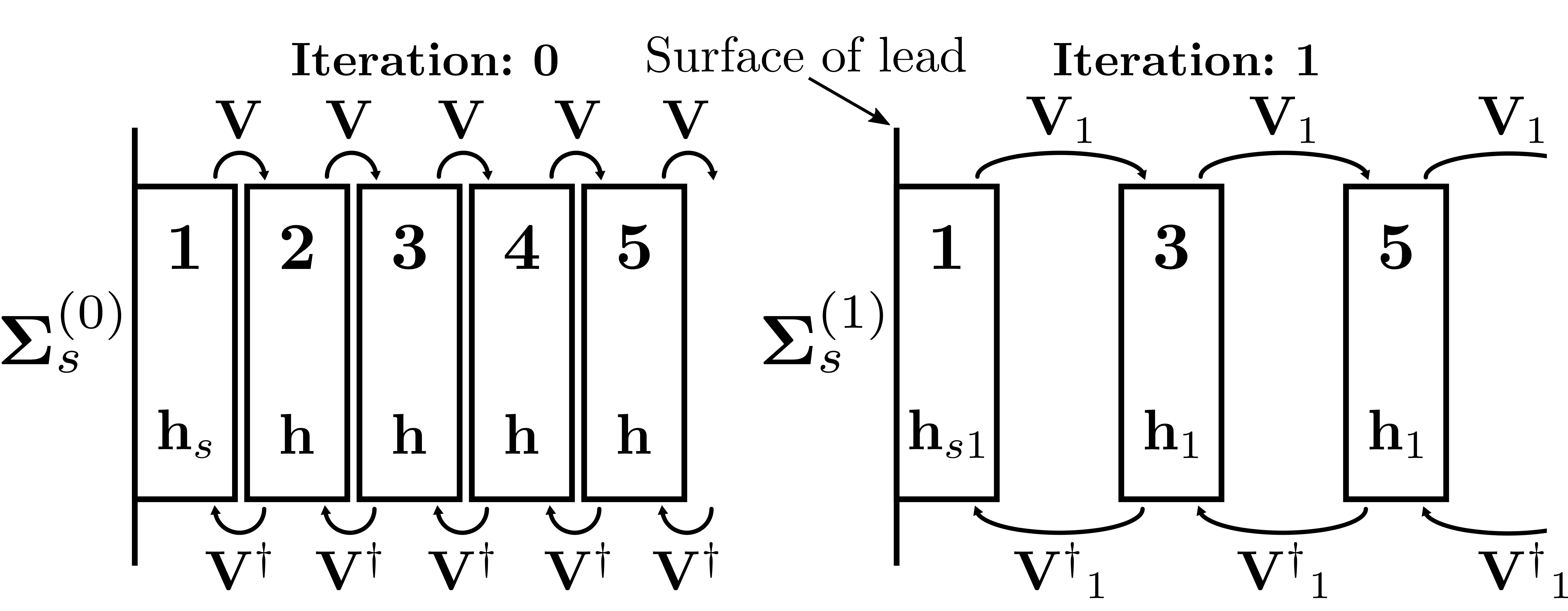}
        \caption{Sketch of how the Rubio-Sancho method operates. Each iteration, $j$, removes every other cell from the lead, updating the values of the local Hamiltonians, $\vec{h}_{sj}$ and $\vec{h}_j$, coupling matrices, $\vec{V}_j^{(\dagger)}$, and surface self energy, $\vec{\Sigma}_{sj}$. The algorithm converges when remaining cells are sufficiently distant such that $\vec{V}_j^{(\dagger)}\rightarrow0$.}
        \label{fig:lead}
    \end{figure}

    The method refactors Eq.~\eqref{eq:dyson_02} for the right semi-infinite, unperturbed (no self energy term) Green's function of the lead into a series of linear equations,
    \begin{eqnarray}
    &&[z\vec{I}-\vec{\mathcal{H}}]\vec{G}_\text{1st}=\\
    &&\left(\begin{array}{c c c c c c}
    z\hat{\vec{I}}-\vec{h}_{s}          & -\vec{V}^\dagger          & 0                         & 0                 & 0         & \cdots \\
    -\vec{V}                            & z\hat{\vec{I}}-\vec{h}    & -\vec{V}^\dagger          & 0                 & 0         & \cdots \\
    0                                   & -\vec{V}                  & z\hat{\vec{I}}-\vec{h}    & -\vec{V}^\dagger  & 0         & \cdots \\
    \vdots                              & \ddots                    & \ddots                    & \ddots            & \ddots    & \ddots
    \end{array}\right)\left(\begin{array}{c c c c}
    1 \\
    0 \\
    0 \\
    \vdots
    \end{array}\right),\nonumber
    \end{eqnarray}
    where $z=\varepsilon+i\zeta$, $\vec{h}$ is the Hamiltonian for each slice in the lead, $\vec{h}_s$ is the Hamiltonian for the slice at the surface of the lead, and $\vec{V}$ are the forward hopping matrices between each slice. In $\vec{G}_\text{1st}=\left(\vec{G}_{00},\vec{G}_{10},\vec{G}_{20},\hdots\right)^T$ we have discarded all but the first column of the full Green's function (which would, in principle, be an infinite square matrix). This is sufficient to calculate the surface Green's function $\vec{G}_{00}$. Using this set of equations we eliminate every other row of the above matrix, resulting in relations,
    \begin{eqnarray}
    \alpha_{j+1}    =&\alpha_j\tilde{g}_j\alpha_j,  \qquad  \chi_{j+1}      =&\chi_j+\alpha_j\tilde{g}_j\beta_j+\beta_j\tilde{g}_j\alpha_j, \nonumber\\
    \beta_{j+1}     =&\beta_j\tilde{g}_j\beta_j,    \qquad  \chi_{s(j+1)}   =&\chi_{s,j}+\alpha_j\tilde{g}_j\beta_j,
    \end{eqnarray}
    where $\chi_{s0}=\vec{h}_s$, $\chi_0=\vec{h}$, $\alpha_0=\vec{V}^\dagger$, $\beta_0=\vec{V}$ and $\tilde{g}_j=(\vec{I}-\chi_j)^{-1}$. As $j$ increases we are coupling slices which are further and further apart, with the effect of the intermediate cells absorbed into those remaining (see Fig.~\ref{fig:lead}).
    Parameters $\alpha_j$ and $\beta_j$ quantify the coupling strength between increasingly distant cells, vanishing as $j\rightarrow\infty$. At this point the entire semi-infinite lead is described by the surface self energy. Numerically, we define a cutoff, $\nu$, after which we consider the iterative process to have converged. Then, the surface Green's function of the lead is given by the summation,
    \begin{equation}
        \vec{G}^{-1}_{00}\approx\vec{g}^{-1}_{00}-\sum_{j=1}^\nu \alpha_j\tilde{g}_j\beta_j,
    \end{equation}
    where $\vec{g}_{00}$ is the unperturbed Green's function of the surface cell and all following terms are self energy corrections. An equivalent expression for the left lead is recovered by exchanging $\alpha_j\leftrightarrow\beta_j$. In our calculations this method converged for $\nu\sim10$ iterations. Then, the self energy of each lead is given by,
    \begin{equation}
        \Sigma_{L} = \vec{V}\vec{G}_{L}\vec{V}^\dagger,\quad\Sigma_{R} = \vec{V}^\dagger\vec{G}_{R}\vec{V},
    \end{equation}
    where $\vec{G}_{L/R}$ is the surface Green's function of the left/right lead.

    \section{\label{C}Further data}

    \begin{figure*}[!htb]
        \centering
        \includegraphics[width=1\linewidth]{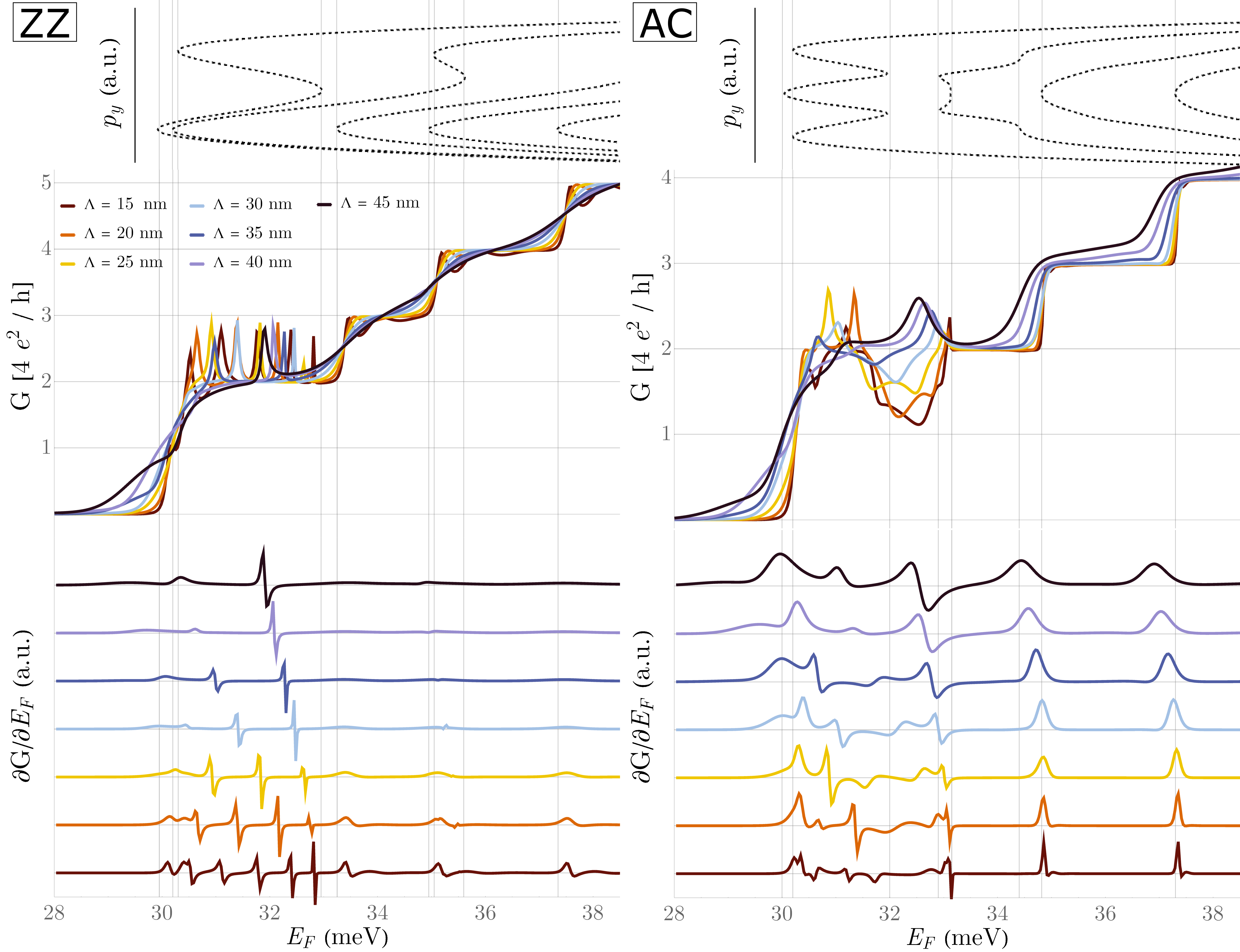}
        \caption{Further conductance data (middle) and differential conductance data (bottom) with corresponding channel spectra (top) for zigzag and armchair lattice orientation along the transport axis (left and right columns respectively) for different values of smoothing length, $\Lambda$. Channel parameters are $U_0=-20$~meV, $\Delta_0=150$~meV, $\beta=0.3$, $\delta=10$~meV, $V_0=-200$~meV, $W=50$~nm and $L=240$~nm.}
        \label{fig:transport}
    \end{figure*}

    In Fig.~\ref{fig:transport} we show further conductance data for a range of lead/channel smoothing distances for a channel with $L=240$~nm. Channels show the coherent forward transmission resonances which enhance the expected adiabatic conductance steps as outlined in Figs.~\ref{fig:transport_zz} and \ref{fig:transport_ac} of the main text. These resonance peaks exhibit a well-defined periodicity related to the length of the semimetallic region of the wire, $\ell$\footnote{Since quantity $L$ in Eq.~3 is defined as the distance between smoothing centres, $\ell<L$ and $\ell\rightarrow0$ as $\Lambda\rightarrow L/2$ (approximately).}, and can be clearly identified in the differential conductance, $\partial G/\partial E_F$ (bottom panel). The shorter maximum channel length (as compared to Figs.~\ref{fig:transport_zz} and \ref{fig:transport_ac}), and correspondingly shorter confining length of the standing waves within the semimetallic region of the wire, results in a larger average energy splitting between consecutive standing wave modes. Therefore, fewer resonance peaks are visible in the first conductance plateau and they are more greatly separated along the energy axis.

    \bibliographystyle{apsrev4-1}
    \bibliography{BLG_channel_PRB_final_2019_bib}

    \end{document}